\documentclass[fleqn,usenatbib]{mnras}

\usepackage{newtxtext,newtxmath}

\usepackage[T1]{fontenc}
\usepackage{ae,aecompl}

\usepackage{graphicx}	
\usepackage{amsmath}	

\title[Dust in the CGM]{Dust enrichment in the circum-galactic medium}

\author[M. Otsuki \& H. Hirashita]{
Mau Otsuki$^{1,2}$\thanks{E-mail: tsukimau@gmail.com} and
Hiroyuki Hirashita$^{2,3}$\thanks{E-mail: hirashita@asiaa.sinica.edu.tw}
\\
$^{1}$Department of Physics, Faculty of Science, Hiroshima University, 1-3-1 Kagamiyama, Higashi-Hiroshima,
Hiroshima 739-8526, Japan\\
$^{2}$Institute of Astronomy and Astrophysics, Academia Sinica,
Astronomy--Mathematics Building, No.\ 1, Section 4,
Roosevelt Road, Taipei 10617, Taiwan\\
$^{3}$Theoretical Astrophysics, Department of Earth and Space Science, Osaka University, 1-1
Machikaneyama, Toyonaka, Osaka 560-0043, Japan
}

\date{Accepted XXX. Received YYY; in original form ZZZ}

\pubyear{2023}

\begin{document}
\label{firstpage}
\pagerange{\pageref{firstpage}--\pageref{lastpage}}
\maketitle

\begin{abstract}
To understand the origin of dust in the circum-galactic medium (CGM),
we develop a dust enrichment model.
We describe each of the central galaxy and its CGM as a single zone,
and consider the mass exchange between them through galactic inflows and outflows.
We calculate the evolution of the gas, metal, and dust masses in the galaxy and the CGM. In the galaxy, we include stellar dust production and interstellar dust processing following our previous models. The dust in the galaxy is transported to the CGM via galactic outflows, and it is further processed by dust destruction (sputtering) in the CGM. 
We parameterize the time-scale or efficiency of each process and investigate the effect on the dust abundance in the CGM. 
We find that the resulting dust mass is sensitive to the dust destruction in the CGM, and the dust supply from galactic outflows, both of which directly regulate the dust abundance in the CGM. The inflow time-scale also affects the dust abundance in the CGM because it determines the gas mass evolution (thus, the star formation history) in the galaxy. The dust abundance in the CGM, however, is insensitive to stellar dust formation in the galaxy at later epochs because the dust production is dominated by dust growth in the interstellar medium. We also find that the resulting dust mass in the CGM is consistent with the value derived from a large sample of SDSS galaxies.
\end{abstract}


\section{Introduction}\label{sec:intro}

Dust is an important component in the interstellar medium (ISM) of
galaxies and its evolution is tightly related to chemical enrichment of galaxies \citep[e.g.][]{Dwek:1998a}.
The existence of dust is not limited to the ISM.
In fact, dust is also found in the intergalactic
medium (IGM) or in the circum-galactic medium (CGM).
Although it is not easy to directly detect dust in the CGM and IGM, large statistical data taken by the Sloan Digital Sky Survey \citep[SDSS;][]{York:2000a} have enabled us to statistically detect dust
in intervening absorption systems towards
quasi-stellar objects (QSOs).
In this paper, we particularly focus on the CGM rather than the IGM because of clearer evidence of dust (see below).

Clarifying the origin and evolution of dust in the CGM is important
because dust has a significant impact on the thermal state of the CGM.
Dust radiates away the thermal energy obtained from collisions with hot gas
if the gas temperature is higher than $\sim 10^6$ K \citep[e.g.][]{Dwek:1987a}.
If the gas temperature is lower,
photoelectric heating by dust could be an important heating mechanism
\citep{Inoue:2003b,Inoue:2004a}.
Therefore, dust in the CGM has a critical importance in understanding how the physical condition of the CGM
has evolved in the cosmic history.

As mentioned above, SDSS data have opened up a way of revealing
the existence of dust in the CGM.
In addition, they have contributed to clarifying the properties of CGM dust.
\citet{York:2006a} succeeded in detecting dust extinction by comparing 
QSOs with and without intervening absorbers. This method also enabled them to derive the typical extinction curve in intergalactic absorbers.
\citet{Menard:2010a} extracted information on reddening (or colour excess) on several-Mpc scales around galaxies at $z\sim 0.3$ (where $z$ is the redshift)
by analyzing the cross-correlation between the galaxies and the reddening of background QSOs.
A similar method was also applied to nearer galaxies at $z\sim 0.05$ by \citet{Peek:2015a}, who  found a radial reddening profile similar to the
one by \citet{Menard:2010a}.
Note that extended dust emissions around galaxies at $z\sim 1$ have recently been suggested by stacking analysis of millimetre data \citep{Meinke:2023a}.
\citet{Masaki:2012a}, using an analytic model,
reproduced the large extent of dust distribution around galaxies.
There are some studies that used Mg \textsc{ii} absorbers, which are considered to trace the CGM environments\citep{Steidel:1994a,Chen:2010a,Nielsen:2013a,Lan:2020a}.
\citet{Menard:2012a} derived reddening at rest-frame ultraviolet wavelengths
in Mg \textsc{ii} absorbers from statistical analysis, and provided useful constraints on dust properties in the CGM.
From the observational data of various dust tracers,
dust in the CGM proves to be important for the cosmic dust budget
\citep{Menard:2010a,Fukugita:2011a}.

Because of the above importance and observations,
we aim to clarify the origin and evolution of dust in the CGM.
The CGM is also the interface between galaxies and the IGM; thus,
clarifying the dust evolution in the CGM leads to the understanding of how the dust produced in galaxies is dispersed in a wide area of the Universe. Since dust and gas are dynamically coupled through the drag force, dust could be supplied to the CGM by galactic outflows
\citep[e.g.][]{Veilleux:2005a}.
Some studies included dust in hydrodynamical simulations and showed that galactic outflows successfully transport the dust to the CGM
\citep{Zu:2011a,McKinnon:2016a,Hou:2017a,Aoyama:2018a}.
Radiation pressure also pushes the interstellar dust towards the circum-galactic space
\citep{Ferrara:1991a,Davies:1998a,Bianchi:2005a,Bekki:2015a,Hirashita:2019b}.
These dust transport mechanisms could account for the presence of dust in the CGM.

Given the importance of the CGM dust, it is useful
to develop an analytical model of dust enrichment in the CGM.
This is complementary to the above hydrodynamical simulations.
\citet{Gjergo:2018} investigated the evolution of dust
in the intracluster medium (although the focus is not exactly laid on the CGM) using hydrodynamical simulations while taking into account dust formation and evolution processes in and around galaxies.
However, complicated nonlinear processes usually make it difficult to pinpoint the most important factor that determines the dust abundance in the CGM. Thus, in this paper, we aim at analytically modelling the dust enrichment processes of the CGM. The central galaxy acts as a factory of dust, which is subsequently injected into the CGM by galactic outflows.
We utilize the dust enrichment model in the galaxy developed in previous studies \citep[e.g.][]{Lisenfeld:1998a,Dwek:1998a,Hirashita:1999a}. This model includes dust production by stars, growth in the ISM, and destruction in supernova (SN) shocks.
The source of dust in the CGM is dusty galactic outflows, while the CGM also acts as a gas reservoir for the galaxy. Thus, outflows and inflows should be considered \citep[e.g.][]{Tumlinson:2017a}.
Dust in the CGM can also be destroyed by sputtering in the hot gas. 
We expect that the analytical model to be developed in this paper will provide a tool easily compared with observations and will serve to identify key processes governing the dust abundance in the CGM.

This paper is organized as follows.
In Section~\ref{sec:model}, we describe the dust evolution model including relevant processes.
In Section~\ref{sec:result}, we show the results and the dependence on the efficiency of each process included in the model.
In Section \ref{sec:discussion}, we provide some extended discussions for parameter dependence, comparison with observations
and prospects for future modelling.
In Section \ref{sec:conclusion}, we give conclusions.

\section{Model}\label{sec:model}

We model the evolution of dust mass in the CGM of a galaxy.
The baryonic content in the galaxy is composed of stars and the ISM,
and we neglect stars in the CGM.
The model is based on the evolution of several baryonic components in the galaxy and the CGM.
We assume that the abundances of metals and dust are homogeneous within each of the galaxy and the CGM.
We neglect the effect of hierarchical build-up of the system, and treat the galaxy and the CGM as two boxes exchanging masses. This simplified approach is useful to extract important processes for dust enrichment in the CGM, giving a basis for future more complicated calculations (e.g.\ cosmological simulations). Below we separately explain the models for the galaxy and the CGM, and we later combine them by considering the mass exchange between these two zones. In this paper, we focus on a `typical' Milky Way-like galaxy, but the resulting masses are simply scaled with the total baryonic mass $M_0$ assumed in the initial condition (Section \ref{subsec:param}).

\subsection{Dust in the galaxy}
%
Since dust enrichment in a galaxy is strongly linked to metal enrichment,
the evolution of dust mass is described based on the chemical evolution model, in which star formation and resulting metal enrichment are consistently modelled \citep[e.g.][]{Lisenfeld:1998a,Dwek:1998a,Zhukovska:2008a}. Below we describe the time evolution of 
the gas mass, $M_\mathrm{g}^\mathrm{gal}$, the metal mass, $M_Z^\mathrm{gal}$, and the dust mass, $M_\mathrm{d}^\mathrm{gal}$ in the galaxy (the superscript `gal' indicates the galaxy).
In our definition, the metals include both gas-phase metals and dust.
The time ($t$) evolution of these three quantities is described by the following set of equations \citep[e.g.][]{Hirashita:1999a,Inoue:2011a}:
\begin{align}
    \frac{\mathrm{d}M_\mathrm{g}^\mathrm{gal}}{\mathrm{d}t} &
    =(-1+\mathcal{R})\psi-O+I,\label{eq:gas_gal}\\
    \frac{\mathrm{d}M_Z^\mathrm{gal}}{\mathrm{d}t} &
    =[Z^\mathrm{gal}(-1+\mathcal{R})+\mathcal{Y}]\psi-Z^\mathrm{gal}O+Z^\mathrm{C}I,\label{eq:metal_gal}\\
    \frac{\mathrm{d}M_\mathrm{d}^\mathrm{gal}}{\mathrm{d}t} &
    =[f_\mathrm{d\star}(Z^\mathrm{gal}\mathcal{R}+\mathcal{Y})-D^\mathrm{gal}]\psi-D^\mathrm{gal}O+D^\mathrm{C}I \nonumber\\
    &\quad+\frac{M_\mathrm{d}^\mathrm{gal}}{\tau_\mathrm{acc}}(1-\delta^\mathrm{gal})
    -\frac{M_\mathrm{d}^\mathrm{gal}}{\tau_\mathrm{SN}},\label{eq:dust_gal}
\end{align}
where $\psi$ is the star formation rate,
$\mathcal{R}$ is the returned fraction of gas from the formed stars, $O$ is the gas outflow rate from the galaxy to the CGM, $I$ is the gas inflow rate from the CGM to the galaxy, $\mathcal{Y}$ is the metal yielded, $f_\mathrm{d\star}$ is the fraction of metals condensed into dust in stellar ejecta, $\tau_\mathrm{acc}$ is the time-scale of dust growth by the accretion of gas-phase metals onto existing dust grains in the ISM (this process is referred to as dust growth), and $\tau_\mathrm{SN}$ is the time-scale of dust destruction by SN shocks in the ISM (referred to as SN destruction). We also define the following three ratios in the galaxy: $Z^\mathrm{gal}\equiv M_Z^\mathrm{gal}/M_\mathrm{g}^\mathrm{gal}$ is the metallicity, $D^\mathrm{gal}\equiv M_\mathrm{d}^\mathrm{gal}/M_\mathrm{g}^\mathrm{gal}$ is the dust-to-gas ratio, and $\delta^\mathrm{gal}\equiv M_\mathrm{d}^\mathrm{gal}/M_Z^\mathrm{gal}=D^\mathrm{gal}/Z^\mathrm{gal}$ is the dust-to-metal ratio. The metallicity and dust-to-gas ratio in the CGM are also defined by replacing the superscript `gal' with `C' (representing `CGM').
We treat $\mathcal{R}$ and $\mathcal{Y}$ as fixed parameters by applying the instantaneous recycling approximation (Section \ref{subsec:param}), but vary $\psi$, $O$, and $I$ according to the gas mass evolution in the galaxy or the CGM (Section \ref{subsec:SF}).
We also calculate the stellar mass $M_\star$ as
$\mathrm{d}M_\star /\mathrm{d}t=(1-\mathcal{R})\psi$.

Equation (\ref{eq:gas_gal}) describes the net gas consumption
by star formation, and the gas loss and gain by the outflow and inflow, respectively.
Equation (\ref{eq:metal_gal}) basically includes the terms similar to those in equation (\ref{eq:gas_gal}) but multiplied by the metallicity of the appropriate zone (the galaxy or the CGM). The increase of the metallicity is also caused by the newly produced metals expressed by the term including $\mathcal{Y}$. Equation (\ref{eq:dust_gal}) contains the dust condensation in stellar ejecta (terms multiplied by $f_\mathrm{d\star}$) and the consumption by star formation ($D^\mathrm{gal}\psi$), which are followed by the outflow and inflow terms (similar to the last two terms in equation \eqref{eq:metal_gal}). We assume that the dust-to-gas ratios of the inflow and outflow are the same as those in the CGM and the galaxy, respectively. Thus, our model assumes complete dynamical coupling between dust and gas. The last two terms describe the interstellar processing of dust regulated by proper time scales: The first is dust growth and the second is dust destruction by SN shocks in the ISM.
The factor $(1-\delta^\mathrm{gal})$ indicates the fraction of metals in the gas phase, which is multiplied to exclude the metals already condensed into dust.

\subsection{Dust in the CGM}

We also construct basic equations for the gas mass, $M_\mathrm{g}^\mathrm{C}$,
the metal mass, $M_Z^\mathrm{C}$, and the dust mass, $M_\mathrm{d}^\mathrm{C}$ in the CGM. In our model, these quantities are governed by the following equations:
\begin{align}
    \frac{\mathrm{d}M_\mathrm{g}^\mathrm{C}}{\mathrm{d}t}&=O-I+(I_\mathrm{IGM}-O_\mathrm{IGM}),\label{eq:gas_CGM}\\
    \frac{\mathrm{d}M_\mathrm{Z}^\mathrm{C}}{\mathrm{d}t}&=
    Z^\mathrm{gal}O-Z^\mathrm{C}I-Z^\mathrm{C}O_\mathrm{IGM},\label{eq:metal_CGM}\\
    \frac{\mathrm{d}M_\mathrm{d}^\mathrm{C}}{\mathrm{d}t}&=
    D^\mathrm{gal}O-D^\mathrm{C}I-\frac{M_\mathrm{d}^\mathrm{C}}{\tau_\mathrm{dest}}-D^\mathrm{C}O_\mathrm{IGM},\label{eq:dust_CGM}
\end{align}
where $\tau_\mathrm{dest}$ is the dust destruction time-scale in the CGM, and the mass inflow from and the outflow to the IGM are also considered and denoted as $I_\mathrm{IGM}$ and $O_\mathrm{IGM}$, respectively.
We also use the dust-to-metal ratio in the CGM, denoted as $\delta^\mathrm{C}(\equiv M^\mathrm{C}_\mathrm{d}/M^\mathrm{C}_Z=D^\mathrm{C}/Z^\mathrm{C})$.

Equation~(\ref{eq:gas_CGM}) describes the net gain of the gas mass
from the galaxy (note that loss/gain for the galaxy is gain/loss for the CGM)
plus the net gas mass obtained from the IGM outside of the CGM.
Equation~(\ref{eq:metal_CGM}) describes the gain/loss of metals
in a similar way to equation~(\ref{eq:gas_CGM}), and assumes that
the metals (and dust) contained in the IGM is negligible. Equation~(\ref{eq:dust_CGM}) also has similar terms to those in equation (\ref{eq:metal_CGM}) but using the dust-to-gas ratios of the appropriate zone (the galaxy or the CGM). This equation also includes the dust destruction term, since the CGM contains hot gas in which the dust is destroyed by sputtering \citep[e.g.][]{Aoyama:2018a}.

The above equations are not closed because of the IGM terms ($I_\mathrm{IGM}$ and $O_\mathrm{IGM}$). Since our framework is only capable of describing the galaxy and the CGM, we adopt a simple assumption to eliminate the IGM terms.
We suppose that the CGM acts as a reservoir of a constant gas mass: 
$\mathrm{d}M_\mathrm{g}^\mathrm{C}/\mathrm{d}t=0$; that is, the right-hand size of equation (\ref{eq:gas_CGM}) is zero. 
In addition, we assume that either $I_\mathrm{IGM}$ or $O_\mathrm{IGM}$ is zero for simplicity, which means that $I_\mathrm{IGM}=I-O$ if $I>O$, and that $O_\mathrm{IGM}=O-I$ if $I\leq O$. 
This leads to the following expression for $O_\mathrm{IGM}$, which appears in equations (\ref{eq:metal_CGM}) and (\ref{eq:dust_CGM}):
\begin{align}
O_\mathrm{IGM}=
\begin{cases}
        0 & \text{if $O-I\leq 0$}, \\
        O-I & \text{if $O-I>0$}. 
\end{cases}
\end{align}
Note that equations (\ref{eq:metal_CGM}) and (\ref{eq:dust_CGM})
do not have terms containing $I_\mathrm{IGM}$ because, as mentioned above,
we neglect dust and metals in the IGM.
With the above settings, equations (\ref{eq:gas_gal})--(\ref{eq:dust_CGM}) can be solved once we give $\psi$, $O$, and $I$ in the next subsection.

\subsection{Star formation, inflow, and outflow}\label{subsec:SF}

Broadly, the evolution described by the above equations is regulated by
$\psi$, $I$, and $O$. We set these quantities based on some physical insights.

We assume that the star formation rate is regulated by the star formation time-scale $\tau_\mathrm{SF}$
\citep[e.g.][]{Asano:2013a}:
\begin{align}
\psi =\frac{M_\mathrm{g}^\mathrm{gal}}{\tau_\mathrm{SF}}.\label{eq:SF}
\end{align}
Star formation occurs in collapsing molecular clouds; thus $\psi$ is basically scaled with the free-fall time-scale in those clouds. In addition, we need to take into account the star formation efficiency determined by stellar feedback (the suppression of star formation due to the energy input from formed stars).
Because stellar feedback is a highly nonlinear process \citep[see][and references therein for a theoretical treatment of stellar feedback in recent simulations]{Oku:2022}, 
the star formation efficiency is uncertain.
Therefore, we treat the star formation time-scale as a parameter.

We also adopt a picture that the stellar feedback drives the mass outflow from the galaxy, so that the following simple proportionality is assumed:
\begin{align}
    O = \eta_\mathrm{out}\psi,
\end{align}
where $\eta_\mathrm{out}$ is the mass-loading factor \citep[e.g.][]{DeVis:2021a}.

We also adopt an infall model in which the galaxy acquires the gas through the gas inflow from the CGM \citep{Tinsley:1980aa}. As often assumed in chemical evolution models \citep{Dwek:1998a,Inoue:2011a}, we adopt the following exponentially decaying infall rate, which expresses efficient infall in the galaxy formation epoch, whose duration is regulated by $\tau_\mathrm{in}$:
\begin{align}
  I = \frac{M_0}{\tau_\mathrm{in}}\exp(-{t}/{\tau_\mathrm{in}}),\label{eq:infall}
\end{align}
where $M_0$ is the total mass of the infall gas (as $t\to\infty$), $\tau_\mathrm{in}$ is the infall time-scale. We practically adjust $M_0$ to reproduce the typical stellar mass of a Milky Way-like galaxy (see Section \ref{subsec:param}). Note that all the masses calculated in the model are scaled with $M_0$.

%
\subsection{Parameters}\label{subsec:param}

We fix some free parameters above and move others that could largely affect the dust mass in the CGM. In what follows, we explain how we determine parameter values and their ranges.

The two time-scales related to dust processing are estimated as follows. The accretion time-scale is inversely proportional to the metallicity and is expressed as \citep{Inoue:2011a}
\begin{align}
\tau_\mathrm{acc}=\tau_\mathrm{acc,0}\left(\frac{Z^\mathrm{gal}}{\mathrm{Z}_{\sun}}\right)^{-1} ,
\end{align}
where $\tau_\mathrm{acc,0}$ is the accretion time-scale at solar metallicity,
and is given as a constant parameter.
The destruction time-scale $\tau_\mathrm{SN}$ is related to $\tau_\mathrm{SF}$ in the following way. According to \citet{McKee:1989a}, $\tau_\mathrm{SN} = M_\mathrm{g}^\mathrm{gal}/(\epsilon_\mathrm{dest}M_\mathrm{s}\gamma_\mathrm{SN})$, where $\epsilon_\mathrm{dest}$ is the destruction efficiency of dust in a single SN blast, $M_\mathrm{s}$ is the mass of the ISM swept by a single SN, and $\gamma_\mathrm{SN}$ is the SN rate. Assuming the instantaneous recycling (zero lifetimes of supernova progenitors), we obtain $\tau_\mathrm{SN}=\tau_\mathrm{SF}/(\epsilon_\mathrm{dest}M_\mathrm{s}\mathcal{F}_\mathrm{SN})$, where $\gamma_\mathrm{SN}\equiv\mathcal{F}_\mathrm{SN}\psi$, using equation~(\ref{eq:SF}).

We basically fix the parameters that only affect the dust enrichment in the galaxy; in particular, the dust growth and destruction have been constrained using the relation between dust-to-gas ratio and metallicity \citep[e.g.][]{Hirashita:2011a,Asano:2013a}.
For the parameters concerning dust destruction,
we adopt $\epsilon_\mathrm{dest}=0.1$ and $M_\mathrm{s}=6800$ M$_{\sun}$ \citep{McKee:1989a,Hirashita:2019}. Using the Chabrier initial mass function (IMF) with a stellar mass range of 0.1--100 M$_{\sun}$, we obtain $\mathcal{F}_\mathrm{SN}=9.9\times 10^{-3}$ M$_{\sun}^{-1}$ \citep{Hirashita:2017a}. Thus,
$\tau_\mathrm{SN}=\tau_\mathrm{SF}/6.8$.
For dust growth, as adopted by \citet{Hirashita:2011a}, $\tau_\mathrm{acc,0}$ is of the order of a few $\times 10^7$ yr. Since not all the ISM hosts dust growth, we conservatively adopt $\tau_\mathrm{acc,0}=0.1$ Gyr.
As we show later, our models reproduce a typical value of dust-to-gas ratio $\sim 0.01$ at $t\sim 10$ Gyr.

The returned fraction ($\mathcal{R}$) and the metal yield ($\mathcal{Y}$) can be determined, given the IMF and the turn-off mass, $m_0$. We adopt the instantaneous recycling approximation; that is, we assume that stars more massive than $m_0$ end their lives instantaneously. Since we are interested in Gyr-time evolution, we adopt $m_0=2$ M$_{\sun}$, of which the corresponding stellar lifetime is $\sim 1$ Gyr.
Using the formulae given in the appendix of \citet{Hirashita:2011a} with the same Chabrier IMF as adopted above, we obtain
$(\mathcal{R}, \mathcal{Y}) = (0.32,\, 2.2\times10^{-2})$ for $m_0=2~\mathrm{M}_\odot$. As shown by \citet{Hirashita:2011a}, $\mathcal{R}$ and $\mathcal{Y} $ are not sensitive to the adopted $m_0$; for example, if we adopt $1~\mathrm{M}_\odot$, we obtain $(\mathcal{R}, \mathcal{Y}) =(0.40,\, 2.2\times10^{-2})$. We also confirmed that the dust mass in the CGM changes only by $\lesssim 10\%$ if we change $m_0$ to 1 M$_{\sun}$. Thus, we hereafter adopt the values for $m_0=2$ M$_{\sun}$.

We move the rest of the parameters:
$\tau_\mathrm{SF}$, $\tau_\mathrm{in}$, $f_\mathrm{d\star}$, $\eta_\mathrm{out}$,
and $\tau_\mathrm{dest}$.
The adopted values and ranges are listed in Table \ref{tab:params}.
The star formation and chemical enrichment of the galaxy are regulated by
$\tau_\mathrm{SF}$ and $\tau_\mathrm{in}$.
The star formation time-scale of Milky Way-like galaxies is a few Gyr \citep[e.g.][]{Asano:2013a}; thus, we employ $\tau_\mathrm{SF}=5$~Gyr for the fiducial value. We adopt the value of $\tau_\mathrm{in}$ comparable to $\tau_\mathrm{SF}$ since the stellar mass build-up is likely linked to the increase of the baryonic mass in the galaxy. Accordingly, we adopt $\tau_\mathrm{in}=5$~Gyr for the fiducial value. We change $\tau_\mathrm{SF}$ and $\tau_\mathrm{in}$ in an order-of-magnitude range of 1--10 Gyr to investigate the effects of the galaxy evolution. For the dust enrichment in the galaxy, we vary $f_\mathrm{d\star}$, which controls the efficiency of stellar dust enrichment. We adopt $f_\mathrm{d\star}=0.1$ for the fiducial value, but vary it considering the uncertainty in the dust yield calculations \citep[0.01--0.5;][and references therein]{Inoue:2011a,Kuo:2013a}. Varying this parameter is also useful to clarify the effect of relative importance between stellar dust production and dust growth in the ISM.
We later find that $\eta_\mathrm{out}$ and $\tau_\mathrm{dest}$ affect the dust mass in the CGM directly. The mass loading factor $\eta_\mathrm{out}$ is not well constrained, but recent simulations indicate that $\eta_\mathrm{out}\sim 2$ in Milky Way-mass halos \citep[e.g.][]{Pillepich:2018a}. Thus, we adopt $\eta_\mathrm{out}=2$ for the fiducial value but vary it in a wide range (0.3--7). If $\eta_\mathrm{out}$ is larger than 7, the galaxy is not able to sustain gas; if it is smaller than 0.3, the dust enrichment in the CGM is too inefficient. For the dust destruction in the CGM, we adopt $\tau_\mathrm{dest}=0.5$--10 Gyr, and choose $\tau_\mathrm{dest}=2$ Gyr for the fiducial value because \citet{Aoyama:2018a} showed, based on a cosmological simulation, that the dust is destroyed in the CGM on Gyr time-scales. 

\begin{table}
\caption{Parameters.}
\begin{center}
\begin{tabular}{lccc}
\hline
Parameter & Fiducial & Minimum & Maximum\\
\hline
$\tau_\mathrm{SF}(\mathrm{Gyr})$
& 5 & 1 & 10 \\
$\tau_\mathrm{in}(\mathrm{Gyr})$
& 5 & 1 & 10  \\
$f_\mathrm{d\star}$  
& 0.1 & 0.01 & 0.5\\
$\eta_\mathrm{out}$
& 2 & 0.3 & 7\\
$\tau_\mathrm{dest}(\mathrm{Gyr})$
& 2 & 0.5 & 10 \\
\hline
\end{tabular}
\label{tab:params}
\end{center}
\end{table}

With the above parameter settings, we solve
equations (\ref{eq:gas_gal})--(\ref{eq:dust_gal})
combined with equations (\ref{eq:metal_CGM}) and (\ref{eq:dust_CGM}).
As noted above, we do not solve equation~(\ref{eq:gas_CGM}) but assume a constant value for $M_\mathrm{g}^\mathrm{C}$.
We adopt $M_0=3\times 10^{11}$ M$_{\sun}$ to achieve a typical stellar mass of a Milky Way-like galaxy ($M_\star\sim 6\times 10^{10}$ M$_{\sun}$; \citealt{Licqua:2015a}). We note again that all the masses at any time are scaled with $M_0$.
For the initial condition, we assume all the masses other than $M_\mathrm{g}^\mathrm{C}$ to be zero ($M_\mathrm{g}^\mathrm{gal}=M_\star =M_Z^\mathrm{gal}=M_\mathrm{d}^\mathrm{gal}=M_Z^\mathrm{C}=M_\mathrm{d}^\mathrm{C}=0$ at $t=0$).
This is based on the infall model \citep{Tinsley:1980aa}, in which the galaxy starts star formation from the gas supplied from the CGM (or the halo).
We choose the fixed value of the CGM mass as $M_\mathrm{g}^\mathrm{C}=M_0/3(=10^{11}~\mathrm{M}_{\sun})$, but the results are not sensitive to this value as long as it is of the order of $M_0$.

\section{Results}\label{sec:result}
\subsection{Effects of the chemical evolution in the galaxy}
\label{sec:fid_result}

\begin{figure*}
    \includegraphics[width=1\linewidth]{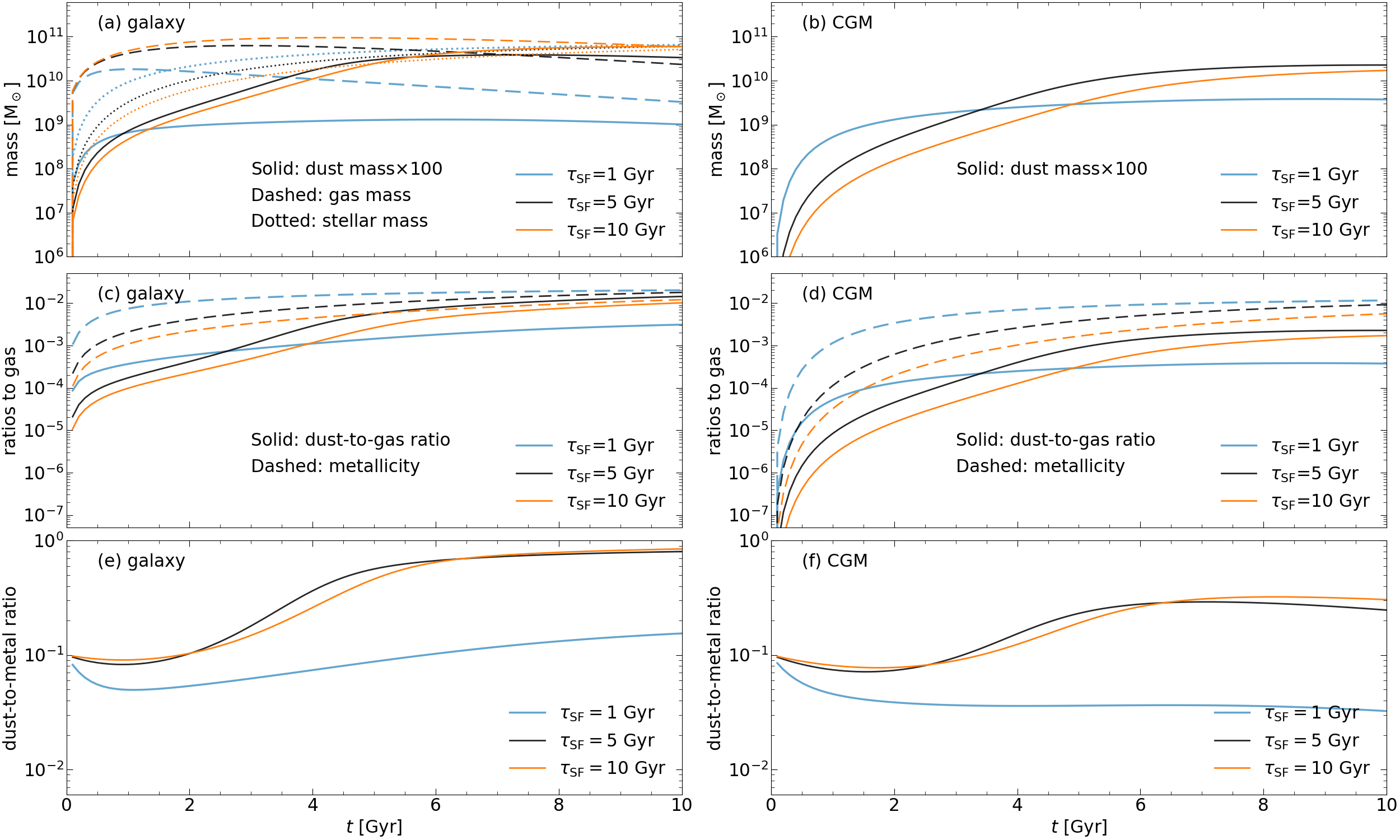}
    \caption{Evolution of (a) the dust (solid), gas (dashed), and stellar (dotted) masses in the galaxy, (b) the dust mass (solid) in the CGM, and (c) the dust-to-gas ratio (solid) and the metallicity (dashed) in the galaxy, (d) those in the CGM, (e) the dust-to-metal ratio in the galaxy, and (f) that in the CGM. Note that the dust mass shown in Panels (a) and (b) is multiplied by 100. The blue, black, and orange lines show the results with $\tau_\mathrm{SF}=1$, 5 (fiducial), and 10 Gyr, respectively. The other parameters are fixed to the fiducial values (Table \ref{tab:params}). We always show the fiducial model with black lines in this and other figures. 
    }
    \label{fig:tSF}
\end{figure*}

\begin{figure*}
    \includegraphics[width=1\linewidth]{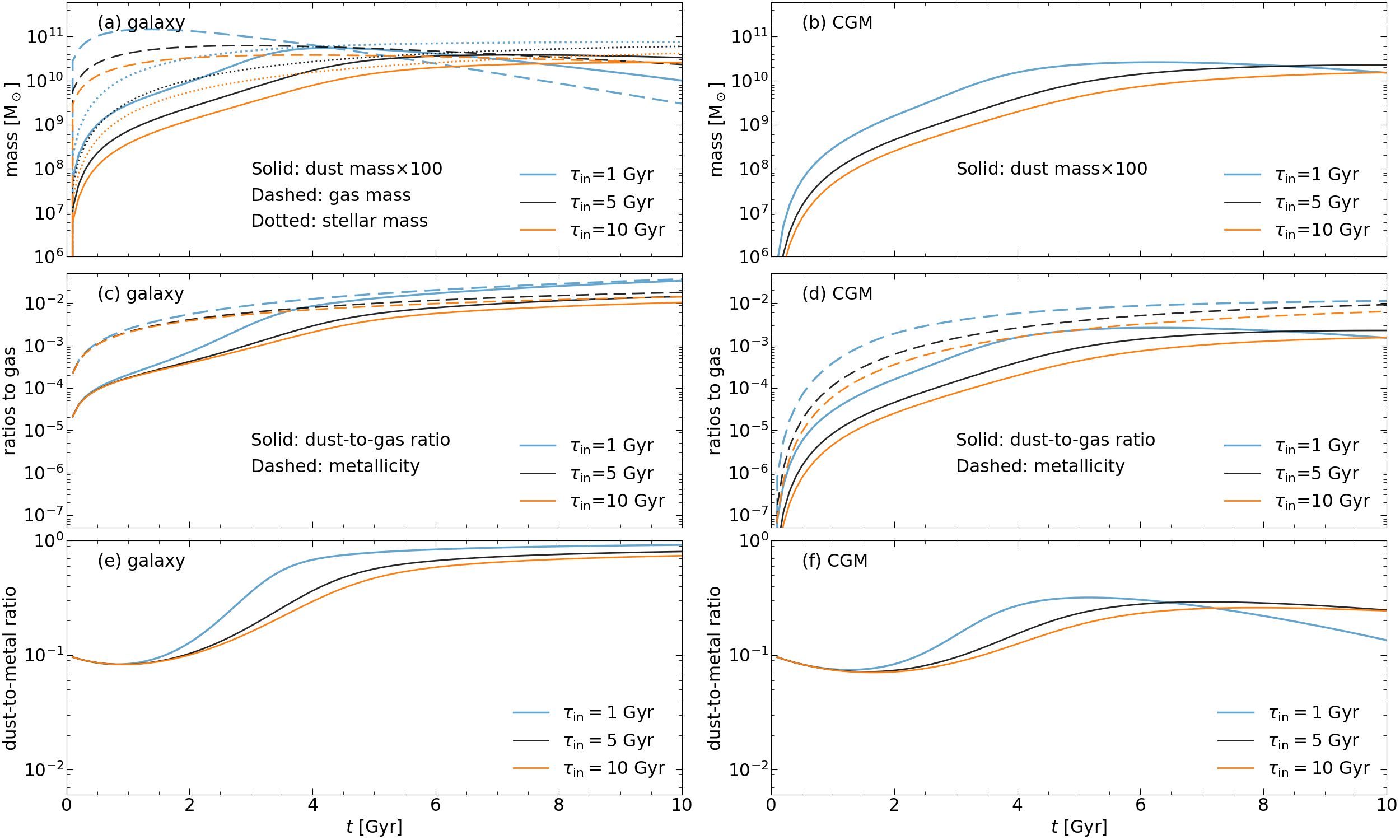}
    \caption{Same as Fig.\ \ref{fig:tSF} but with various $\tau_\mathrm{in}$. The blue, black, and orange lines show $\tau_\mathrm{in}=1$, 5, and 10 Gyr, respectively.}
    \label{fig:tINF}
\end{figure*}

Since the dust production in the galaxy is the source of the CGM dust,
we first examine the effect of the dust enrichment in the galaxy.
Here we focus on the chemical evolution, which is mainly regulated by $\tau_\mathrm{SF}$ and $\tau_\mathrm{in}$ as mentioned in Section \ref{subsec:param}. We show the results for various values of $\tau_\mathrm{SF}$ and $\tau_\mathrm{in}$ in Figs.\ \ref{fig:tSF} and \ref{fig:tINF}, respectively. Other than the varied parameter for each case, we fix the values to the fiducial ones (Table \ref{tab:params}).
In the figures, we present the evolution of the dust masses
in the galaxy and in the CGM. In addition, we present the gas mass, which is useful to see the effect of inflow and gas consumption, as well as the stellar mass in the galaxy. We also display the dust-to-gas ratio, metallicity, and dust-to-metal ratio: Since these three quantities are not affected by the total available baryonic mass (i.e.\ $M_0$), they are more robustly predicted.

\subsubsection{Fiducial case}

We first describe the results in the fiducial case
(see the case of $\tau_\mathrm{SF}=5$ Gyr in Fig.\ \ref{fig:tSF}).
The gas mass increases in the beginning because of the inflow while it gradually declines at later stages because of gas consumption by star formation and outflow. The metallicity continues to increase as a result of enrichment from formed stars. A part of the metals are injected into the CGM, enriching the CGM with metals. As a consequence, metal enrichment occurs coherently both in the galaxy and in the CGM. 

The dust mass in the galaxy first grows by the stellar dust production at $t\lesssim 3$ Gyr.
The increase of dust mass is slightly accelerated at $t\sim 3$ Gyr because of dust growth.
Dust grows up to a point where the dust-to-gas ratio becomes comparable to the metallicity in the galaxy.
Note that dust growth is saturated after that because of the decrease in available gas-phase metals.
The dust-to-metal ratio decreases in the early stage because of dust destruction by SN shocks,
while it starts to grow later due to dust growth. 
After $t\sim 6$ Gyr, the dust-to-metal ratio in the galaxy is almost constant, which means that dust growth is saturated. At this late epoch, the dust mass decreases together with the gas consumption and the outflow.

The dust mass in the CGM broadly traces the dust enrichment in the galaxy,
but there are some features worth mentioning.
The dust mass is almost constant at later epochs because of the balance between dust supply from the galaxy and dust destruction in the CGM. In the end, the dust mass in the CGM becomes comparable to that in the galaxy. 
The evolution of the dust-to-gas ratio traces that of the dust mass, but the dust-to-gas ratio in the CGM is several times smaller than that in the galaxy. This reflects the difference in metallicity.
The dust-to-metal ratio in the CGM follows that in the galaxy; however, it does not increase as much as in the galaxy because of dust destruction in the CGM. The decrease of dust-to-metal ratio is clear at $ t\sim6$ Gyr in the CGM. 

\subsubsection{Dependence on $\tau_\mathrm{SF}$}

Now we examine the dependence on the star formation time-scale $\tau_\mathrm{SF}$ (Fig.\ \ref{fig:tSF}).
Broadly, higher dust abundances can be achieved in the CGM if the galaxy is enriched with dust more efficiently.
In what follows, we describe more closely the results.

For the case with short $\tau_\mathrm{SF}$ (1 Gyr), the gas mass in the galaxy starts to decrease already at $t\sim 1$ Gyr because of efficient gas consumption and outflows. In this case, the metal enrichment also proceeds quickly in the galaxy. The dust-to-gas ratio in the galaxy is also higher up to $t\sim 3$ Gyr compared with the other cases with longer $\tau_\mathrm{SF}$, but at later epochs the increase of dust-to-gas ratio slows down because of efficient SN destruction. Recall that $\tau_\mathrm{SN}$ is proportional to $\tau_\mathrm{SF}$ (Section \ref{subsec:param}). Around the solar metallicity, which is achieved after $t\sim 3$ Gyr, the dust growth time-scale is of the order of 0.1 Gyr, which is comparable to the dust destruction time-scale ($\tau_\mathrm{SN}=\tau_\mathrm{SF}/6.8\sim 0.15$ Gyr for $\tau_\mathrm{SF}=1$ Gyr).
Accordingly, the dust-to-metal ratios in the galaxy and the CGM stay at low values in the case of $\tau_\mathrm{SF}=1$ Gyr.

The dust abundance for longer $\tau_\mathrm{SF}$ (5 and 10 Gyr) is much enhanced compared with
the case with $\tau_\mathrm{SF}=1$ Gyr both in the galaxy and in the CGM
because dust growth, of which the time-scale is much shorter than that of SN destruction, efficiently increases the dust mass.
Although the metallicity in the CGM is lower for longer $\tau_\mathrm{SF}$ because of less efficient outflow, the dust mass and dust-to-gas and dust-to-metal ratios in the CGM do not follow the order of metallicity.
In other words, the dust abundance in the CGM is not a monotonic function of $\tau_\mathrm{SF}$. The dust mass and dust-to-gas ratio in the CGM are higher in the fiducial ($\tau_\mathrm{SF}=5$ Gyr) case than in the cases with shorter and longer $\tau_\mathrm{SF}$. If $\tau_\mathrm{SF}$ is as long as 10 Gyr, the dust enrichment proceeds less quickly than in the fiducial case, and at the same time, the outflow (i.e.\ CGM enrichment) is less efficient. Thus, there is an `optimum' $\tau_\mathrm{SF}$ for dust enrichment in the CGM; that is, the CGM is enriched with dust the most efficiently if $\tau_\mathrm{SF}$ is long enough for SN destruction to occur slowly but is still short enough for outflows and dust enrichment to occur quickly.

\subsubsection{Dependence on $\tau_\mathrm{in}$}

Here we vary the inflow time-scale $\tau_\mathrm{in}$ (Fig.\ \ref{fig:tINF}).
At $t\lesssim 3$ Gyr, the galaxy is supplied with more gas for shorter $\tau_\mathrm{in}$,
which triggers rapid star formation leading to more dust and metals.
Accordingly, the metallicity becomes higher for shorter $\tau_\mathrm{in}$, making dust growth active. Thus, the dust-to-gas ratio is higher for shorter $\tau_\mathrm{in}$ in the galaxy. Shorter $\tau_\mathrm{in}$ also means less gas supply at later epochs so that the dust mass together with the gas mass declines at later epochs in the galaxy.
As a consequence, the dust mass in the galaxy becomes the smallest in the case of $\tau_\mathrm{in}=1$ Gyr after $t\sim$ 8 Gyr. In the case of $\tau_\mathrm{in}=10$ Gyr, the dust enrichment proceeds more slowly than the two other cases as expected.

The enrichment of dust and metals in the CGM basically follows
that in the galaxy, but there are some features specific for the CGM.
In the case of short $\tau_\mathrm{in}(=1~\mathrm{Gyr})$, the dust supply from the galaxy to the CGM declines significantly at later epochs because of the decrease of star formation activity. Thus, the effect of dust destruction becomes prominent and decreases the dust abundance in the CGM.
In the cases of longer $\tau_\mathrm{in}$ (5 and 10 Gyr), the dust mass and the dust-to-gas ratio in the CGM monotonically increase. The dust mass and dust-to-gas ratio at $t=10$ Gyr are not monotonic functions of $\tau_\mathrm{in}$ for the same reason as for the dependence on $\tau_\mathrm{SF}$; there is an `optimum' $\tau_\mathrm{in}$ for dust enrichment in the CGM.

The dust-to-metal ratio is the highest in the case of the shortest
$\tau_\mathrm{in}(=1~\mathrm{Gyr})$ in the galaxy,
reflecting a high efficiency of dust growth due to high metallicity. 
In the CGM, the dust-to-metal ratio increases at $t\sim 3$--5 Gyr
because of dust growth in the galaxy as seen in the synchronous
increase of the dust-to-metal ratio in the galaxy.
The dust-to-metal ratio in the CGM decreases at later epochs,
especially for short $\tau_\mathrm{in}$ because dust destruction exceeds dust supply.
In the case of long $\tau_\mathrm{in}$ (= 10 Gyr), the dust-to-metal ratio grows more slowly, although it reaches almost the same value at $t=10$ Gyr as in the fiducial case.

\subsection{Effects of stellar dust production}
%
\begin{figure*}
    \centering
    \includegraphics[width=1\linewidth]{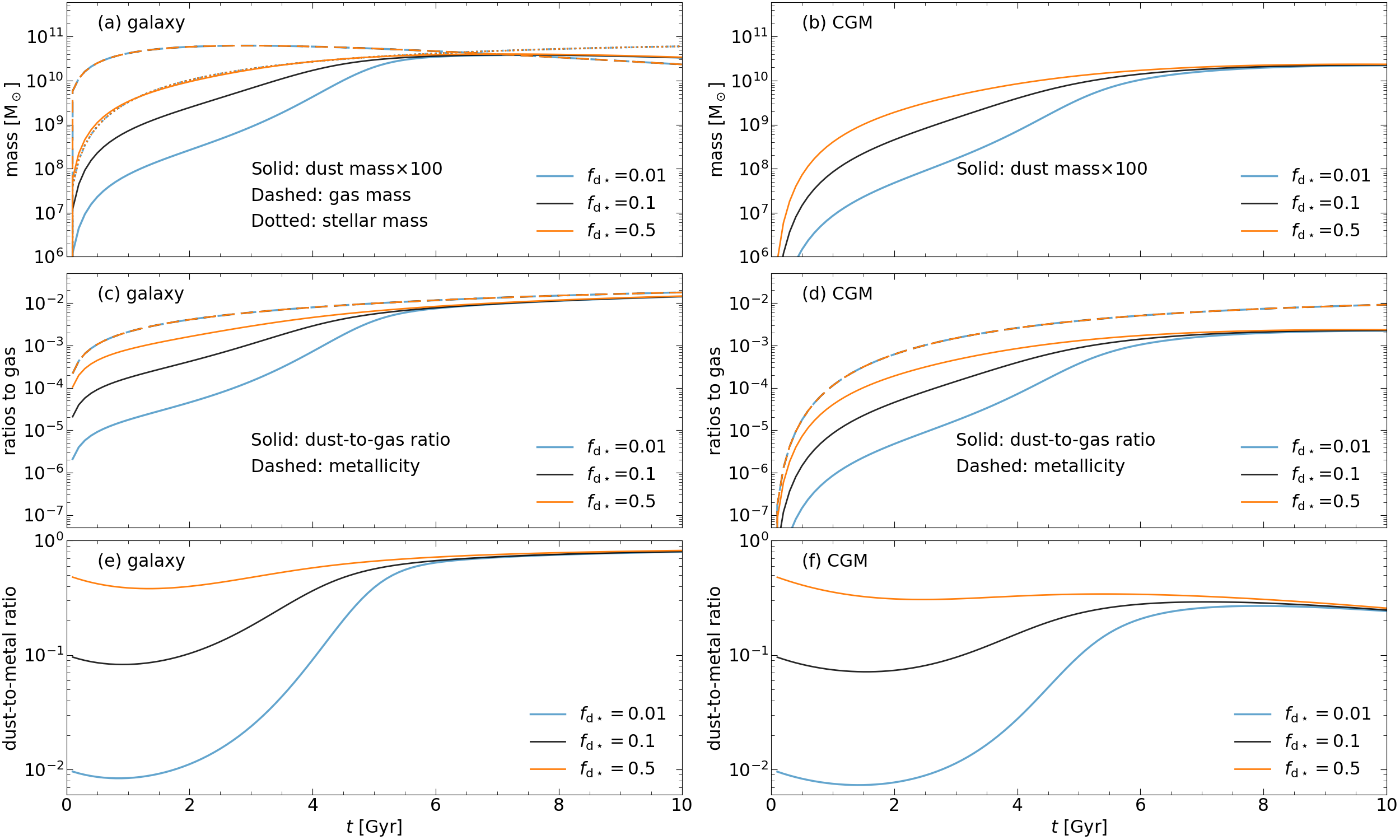}
    \caption{Same as Fig.\ \ref{fig:tSF} but with various values of $f_\mathrm{d\star}$.
    The blue, black, and orange lines show $f_\mathrm{d\star}=0.01$, 0.1, and 0.5, respectively. Note that the three cases for $f_\mathrm{d\star}$ overlap for
    the gas mass (Panel a) and the metallicity (Panels c and d) because these quantities are not affected by stellar dust production.}
    \label{fig:fDstar}
\end{figure*}

Here we investigate the effects of stellar dust enrichment efficiency by varying $f_\mathrm{d\star}$. This parameter could directly regulate the dust abundance, but its value is uncertain (Section \ref{subsec:param}). Thus, we investigate how much $f_\mathrm{d\star}$ (or the uncertainty in this parameter) affects the resulting dust enrichment in the CGM.

We show the evolution of relevant quantities in Fig.\ \ref{fig:fDstar}.
Naturally, $f_\mathrm{d\star}$ does not affect gas or metals.
We find that $f_\mathrm{d\star}$ has a large influence on the dust mass, the dust-to-gas ratio, and the dust-to-metal ratio at $t\lesssim 5$ Gyr.
This time is roughly determined by $\tau_\mathrm{SF}$, which is interpreted as a metal-enrichment
time-scale by formed stars.
As expected, the dust mass is almost proportional to $f_\mathrm{d\star}$ in the early stage. At later epochs, the dust abundance is not sensitive to $f_\mathrm{d\star}$ in the galaxy or the CGM because the main source of dust is dust growth, whose efficiency is independent of $f_\mathrm{d\star}$. 

%
\subsection{Effects of dust supply and destruction in the CGM}
%
Finally, we investigate the effects of two parameters which directly influence the dust mass in the CGM: the mass-loading factor $\eta_\mathrm{out}$ and the dust destruction time-scale $\tau_\mathrm{dest}$.

\subsubsection{Dependence on $\eta_\mathrm{out}$}

\begin{figure*}
    \centering
    \includegraphics[width=1\linewidth]{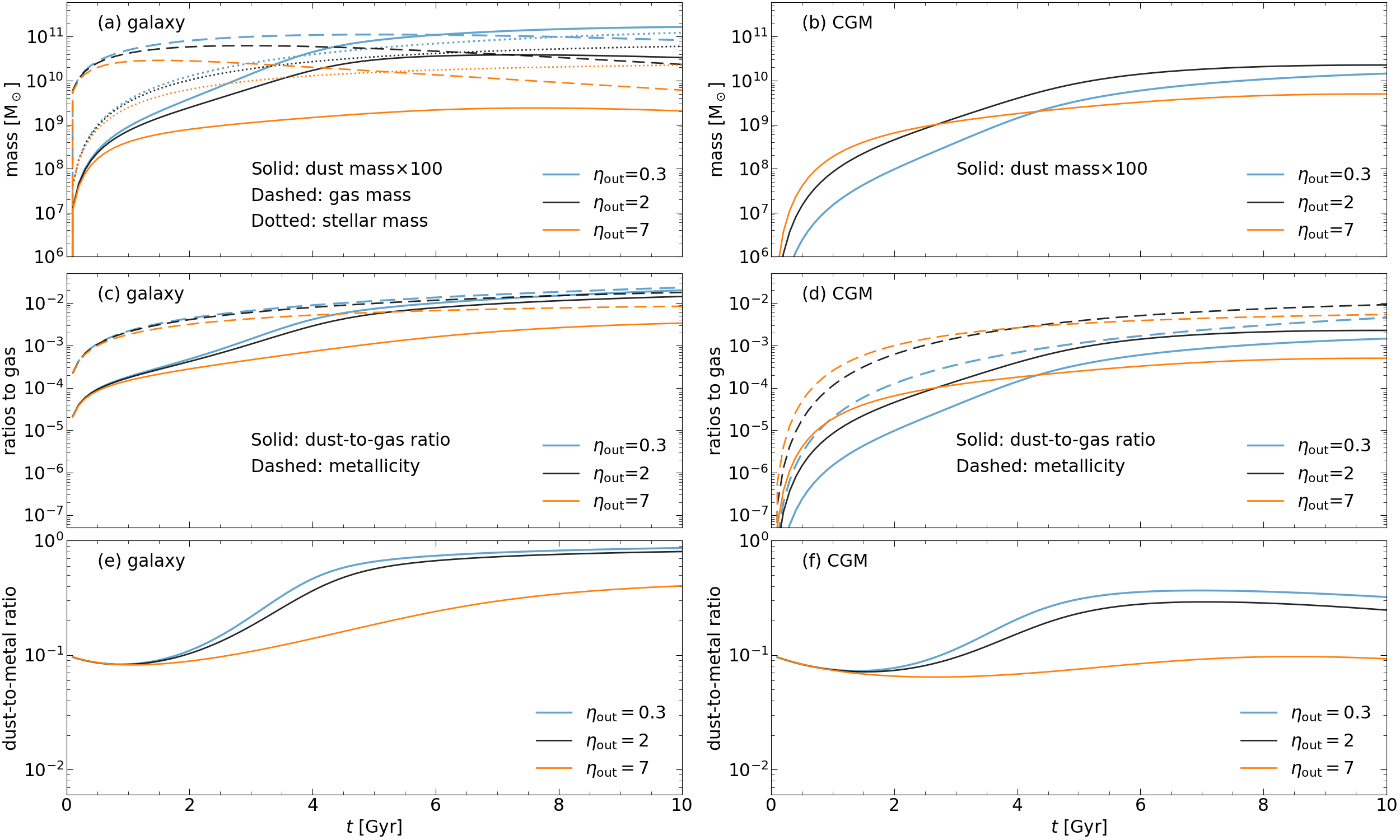}
    \caption{Same as Fig.\ \ref{fig:tSF} but with various $\eta_\mathrm{out}$. The blue, black, and orange lines show $\eta_\mathrm{out}=0.3$, 2, and 7, respectively.}
    \label{fig:mlf}
\end{figure*}

We first examine the effects of varying $\eta_\mathrm{out}$ (Fig.\ \ref{fig:mlf}).
In the case with large $\eta_\mathrm{out}(=7)$, the galaxy quickly loses gas
because of the efficient outflow.
This prevents the galaxy from sustaining star formation, and the galaxy
becomes poor in metals and dust. The low metal content leads to inefficient
dust growth. As a result, the dust abundances remain low both in the galaxy and in the CGM.
For smaller values of $\eta_\mathrm{out}(=2$ and 0.3), in contrast, the galaxy retains dust/metal-enriched gas. 
For this reason, the dust mass, dust-to-gas ratio, and dust-to-metal ratio in the galaxy is higher
if $\eta_\mathrm{out}$ is smaller.

As for the CGM, the galaxy supplies the CGM with more dust mass in the case of larger $\eta_\mathrm{out}$ in the early stage because of the high efficiency of outflow. At later epochs, however, the case of $\eta_\mathrm{out}=7$ predicts less dust mass than the other cases since the star formation (and outflow) declines faster and dust growth is less efficient in the galaxy. This leads to the smallest dust mass and dust-to-gas ratio in the CGM for $\eta_\mathrm{out}=7$. 
In the case of $\eta_\mathrm{out}=0.3$, the outflow is not so efficient as in the fiducial case, which leads to less dust enrichment in the CGM, even though the galaxy contains more dust. From the above comparisons among various values of $\eta_\mathrm{out}$, we observe that there is an `optimum' $\eta_\mathrm{out}$ for dust enrichment in the CGM in the sense that too large or too small a value of $\eta_\mathrm{out}$ leads to less dust content in the CGM.

The dust-to-metal ratio in the CGM reflects that in the galaxy; that is, it is the highest for $\eta_\mathrm{out}=0.3$ and the lowest for $\eta_\mathrm{out}=7$.
The dust in the CGM is decreased by sputtering, whose effect is further investigated below.

\subsubsection{Dependence on $\tau_\mathrm{dest}$}
\begin{figure*}
    \centering
    \includegraphics[width=1\linewidth]{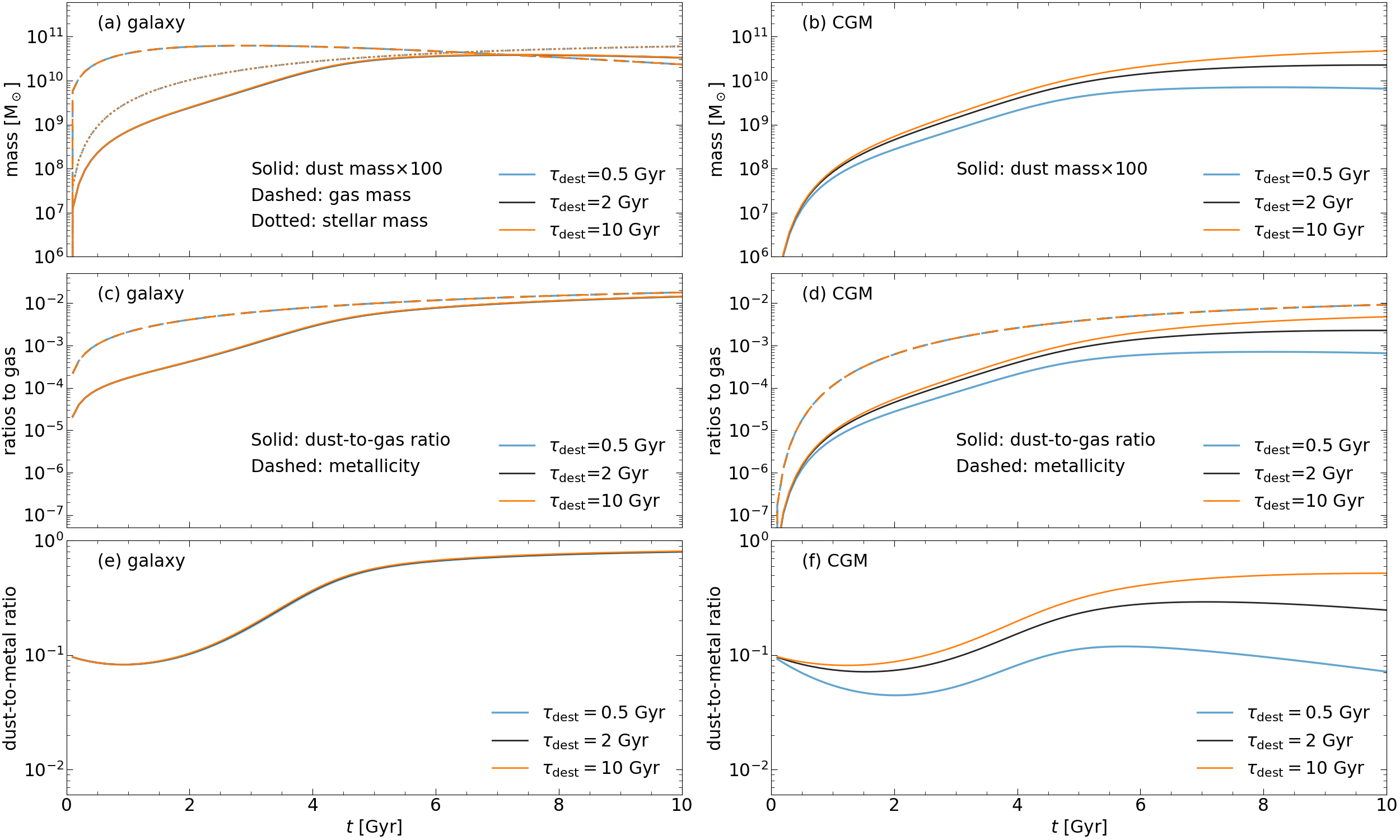}
    \caption{Same as Fig.\ \ref{fig:tSF} but for various values of $\tau_\mathrm{dest}$.
    The blue, black, and orange lines show $\tau_\mathrm{dest}=0.5$, 2, and 10, respectively.
    Since $\tau_\mathrm{dest}$ directly affects dust evolution in the CGM and does not influence other components, lines that describe the evolution in the galaxy (Panels a and c) and metallicity (Panel d) almost overlap.
    }
    \label{fig:tDEST}
\end{figure*}

Finally, we vary $\tau_\mathrm{dest}$ and show the results in Fig.\ \ref{fig:tDEST}.
We confirm that this parameter does not affect the evolution of the galaxy.
As expected, the dust mass in the CGM is smaller for shorter $\tau_\mathrm{dest}$. The increasing rates of dust mass and dust-to-gas ratio become $\sim 0$ at
$t\sim 10$ Gyr for $\tau_\mathrm{dest}=0.5$ and 2 Gyr,
while they are still positive up to $t\sim 10$ Gyr for $\tau_\mathrm{dest}=10$ Gyr.
Note that the metallicity evolution is not affected by $\tau_\mathrm{dest}$.

If we apply smaller values of $\tau_\mathrm{dest}$ (0.5 and 2 Gyr),
the dust-to-metal ratio starts to decline at $t\sim5$ Gyr, when the dust destruction
becomes dominant over the dust supply from the galaxy.
This is because the outflow (i.e.\ dust supply to the CGM) diminishes after $t\sim\tau_\mathrm{SF}=5$ Gyr.
In the case of long $\tau_\mathrm{dest}=10$ Gyr, the dust-to-gas ratio continues to increase even after $t\sim5$ Gyr and achieves the highest value at $t\sim10$ Gyr among the three cases shown.

\section{Discussion}\label{sec:discussion}
%
%

\subsection{Summary of the parameter dependences}
\begin{figure*}
    \centering
    \includegraphics[width=1\linewidth]{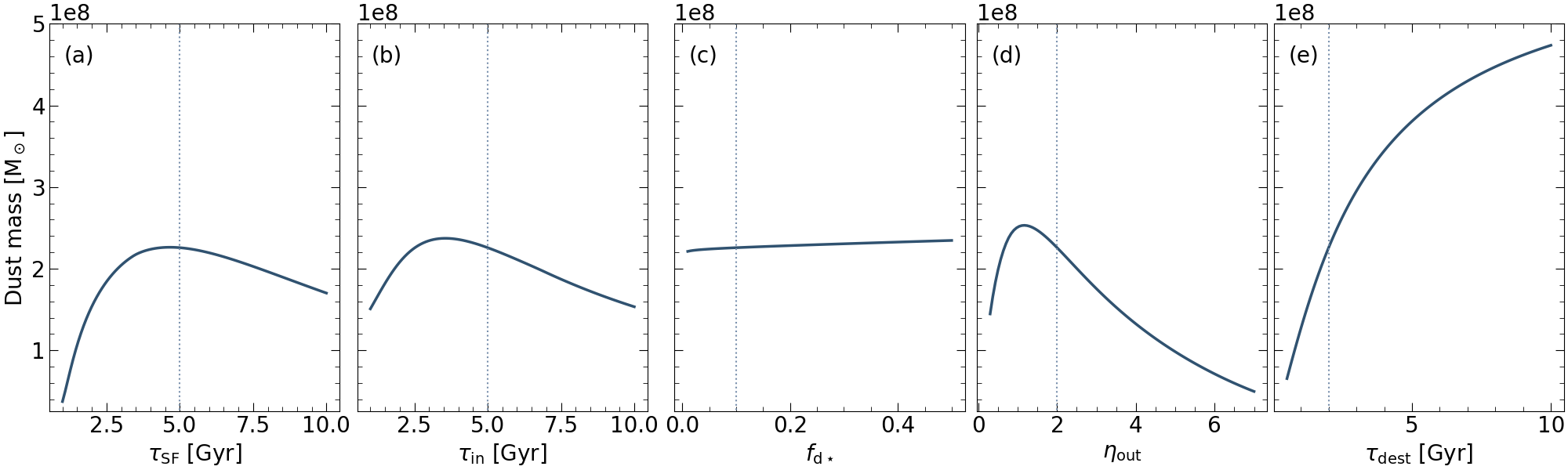}
    \caption{
    Dust mass in the CGM at $t=10$ Gyr as a function of (a) $\tau_\mathrm{SF}$, (b) $\tau_\mathrm{in}$, (c) $f_\mathrm{d\star}$, (d) $\eta_\mathrm{out}$, and (e) $\tau_\mathrm{dest}$ in the range between the maximum and minimum values listed in Table \ref{tab:params}. The dotted vertical line shows the fiducial value of each parameter.}
    \label{fig:cMd_at_10}
\end{figure*}

In the above, we considered three discrete values for each parameter listed
in Table \ref{tab:params}.
Although the effect of each process can be roughly understood
by the three representative values, it is still useful to show
the continuous dependence on each parameter.
We focus on ages comparable to the present cosmic age because of the availability of observational data. Thus, we adopt the calculation results at $t=10$ Gyr, which is used as a typical age of present-day galaxies, and show the continuous dependence on
each parameter in Fig.\ \ref{fig:cMd_at_10}.

From Fig.\ \ref{fig:cMd_at_10} and the result in Section \ref{sec:result}, we find that three parameters play an important role in governing the dust abundance in the CGM at $t \sim10$ Gyr: $\tau_\mathrm{SF}$, $\eta_\mathrm{out}$, and $\tau_\mathrm{dest}$. These three strongly influence the dust evolution in the CGM through chemical evolution in the galaxy, dust supply to the CGM, and dust destruction in the CGM. 

The parameters involved in chemical enrichment and dust supply to the CGM
($\tau_\mathrm{SF}$ and $\eta_\mathrm{out}$, respectively)
do not monotonically change the resulting dust mass in the CGM
(Figs.\ \ref{fig:cMd_at_10}a and d).
The dust mass in the CGM at $t=10$ Gyr is maximized at certain values of
these parameters.
If $\tau_\mathrm{SF}$ is short or $\eta_\mathrm{out}$ is large, dust mass increase in the CGM is suppressed
because of the deficiency of dust formation in the galaxy.
Large $\eta_\mathrm{out}$ also depletes the gas in an early stage so that the metallicity of the galaxy remains low.
In the opposite extremes (long $\tau_\mathrm{SF}$ and small $\eta_\mathrm{out}$), the transport of dust to the CGM is inefficient and the dust mass in the CGM at $t=10$ Gyr could be reduced significantly.
Thus, extreme values of $\tau_\mathrm{SF}$ and $\eta_\mathrm{out}$ lead to inefficient dust enrichment in the CGM, and
the dust mass in the CGM takes a maximum value in the middle of the ranges.
In other words,
there is an optimum choices for the values of $\tau_\mathrm{SF}$ and $\eta_\mathrm{out}$ that maximize the dust mass in the CGM.

The resulting dust mass in the CGM is a monotonic function of $\tau_\mathrm{dest}$ (Fig.\ \ref{fig:cMd_at_10}e).
This is because of the straightforward nature of the parameter $\tau_\mathrm{dest}$:
long/short $\tau_\mathrm{dest}$ simply means that the dust is destroyed inefficiently/efficiently.

Although the effect is smaller compared to the above three parameters,
$\tau_\mathrm{in}$ also indirectly affects the chemical evolution
in the galaxy.
The inflow regulates the amount of gas the galaxy acquires,
subsequently causing star formation, which contributes to
metal/dust enrichment in the galaxy.
The dust is further injected into the CGM through the outflow.
Therefore, $\tau_\mathrm{in}$ effectively plays a similar role to $\tau_\mathrm{SF}$
and $\eta_\mathrm{out}$. This similarity is also clear from Fig.\ \ref{fig:cMd_at_10}:
$\tau_\mathrm{in}$ does not monotonically affect the resulting dust mass
in the CGM (Fig.\ \ref{fig:cMd_at_10}b). When the galaxy acquires gas rapidly with short $\tau_\mathrm{in}$, star formation activity is quenched earlier because of the rapid decline of the inflow. Long $\tau_\mathrm{in}$ makes the whole evolution process slow, and consequently the dust mass in the CGM increases slowly. Thus, there is an `optimum' value of $\tau_\mathrm{in}$ that maximizes the CGM dust mass. 

The stellar dust production regulated by $f_\mathrm{d\star}$ has little
impact on the dust mass in the CGM at $t=10$ Gyr
(Fig. \ref{fig:cMd_at_10}c). This is because
dust growth, not stellar dust production, is the dominant source of dust at later epochs.

\subsection{Comparison with observations}

\citet{Menard:2010a} estimated dust mass in the CGM to be $\sim 5\times10^7~\mathrm{M}_\odot$ at $z\sim 0.3$.
Although the cosmic age at $z\sim 0.3$ is younger than 10 Gyr,
we still use the results at $t=10$ Gyr for comparison with the observation
since it is not meaningful to fine-tune the age and parameters. Moreover, the
dust abundance broadly has a flat age dependence at later ages according to the
results in Section \ref{sec:result}.

\citet{Menard:2010a}'s sample has a mean stellar luminosity
of $\simeq 0.45L^*$, where $L^*$ is a characteristic luminosity
in the luminosity function.
This luminosity corresponds to
$\sim 0.45\times10^{10}~\mathrm{L_\odot}$, using $M^*_{B}=-19.5$ \citep[$B$-band magnitude of the $L^*$ galaxies;][]{Menard:2010a} and the
a solar $B$-band absolute magnitude of 5.48 \citep{Scott:2000a}.
Thus, if the solar mass-to-light ratio is assumed, the mean stellar mass in
the observational sample is estimated as $\sim 0.45\times 10^{10}$~M$_\odot$. 
Our fiducial result aimed at reproducing a Milky Way-like galaxy gives a dust mass of
$M_\mathrm{d}\sim 2\times10^8~\mathrm{M}_\odot$ at $t=10$ Gyr and a stellar mass of
$M_\star \sim 6\times10^{10}~\mathrm{M_\odot}$ (Section \ref{subsec:param}; Fig.\ \ref{fig:tSF}).
Thus, our fiducial galaxies are 6/0.45 times more massive than the above observational sample in terms of the stellar mass conserving the dust-to-stellar mass ratio $\sim 3\times 10^{-3}$. Because of this scaling, we expect that the
$0.45L^*$ galaxies have a dust mass of $2\times 10^8\times 0.45/6=1.5\times 10^7$ M$_\odot$, which is roughly consistent with the dust mass in the observational sample within a factor of $\sim$3.
Considering the uncertainties in the dust mass in \citet{Menard:2010a}
due to e.g.\ the assumed dust mass absorption coefficient, we argue that our model is successful in reproducing the dust mass in the CGM.

\citet{Meinke:2023a} have recently performed stacking analysis for millimetre data of galaxies at $z\sim 1$. Their result is indicative of extended dust emission around galaxies. The typical stellar mass in their sample is larger than the Milky Way value, but the dust-to-stellar mass ratio can be used for comparison. In their data, it is difficult to isolate the component associated with the CGM; however, we could tentatively assume that half of the dust mass is associated with the CGM as is consistent with our fiducial model. In this case, their result suggests that the dust-to-stellar mass ratio for the CGM dust is $\sim 10^{-3}$, which is consistent with our fiducial value within a factor of 3. This together with the above comparison confirms that our dust enrichment scenario of the CGM can explain the mass of the extended dust components observed with various dust tracers.

\subsection{Uncertainties and future prospects}

In the model developed in this paper,
we applied some simplifications for analytical convenience.
We neglected spatial distributions and assumed instantaneous mixing
within each of the galaxy and the CGM.
However, as shown by \citet{Hu:2019a}, the dust-to-gas ratio may be inhomogeneous in the ISM because of dust destruction by SNe.
Since SNe could induce galactic outflows,
the dust-to-gas ratio and the dust-to-metal ratio in the outflows could be different from the mean values in the ISM.
This inhomogeneity needs to be checked by hydrodynamical simulations.

Further physical processes may be worth investigating.
As for the mechanisms of transporting dust from the galaxy to the CGM,
we only considered galactic outflows under an assumption of dynamical coupling between dust and gas.
In this framework, we did not include radiation pressure, which is also suggested to be
a possible mechanism of injecting dust into the CGM
\citep{Ferrara:1991a,Bianchi:2005a,Hirashita:2019b}.
This additional mechanism of dust transport would make the dust enrichment in the CGM more efficient.
However, dust--gas decoupling, which strongly depends on the grain radius, should be taken into account to correctly
treat dust transport by radiation pressure
\citep[e.g.][]{Hirashita:2019b}.
Although some simulations have focused on the evolution of the CGM \citep[e.g.][]{Davies:2020},
it is still challenging to include detailed dust--gas dynamics (including the above decoupling effect) in hydrodynamical simulations because
of requirements for achieving high spatial resolutions and for modelling grain size distributions.
Hydrodynamical simulations are also important since the sputtering efficiency strongly depends on the gas temperature and density.

We only considered the total dust mass, and did not explicitly treat
grain compositions or grain size distributions.
\citet{Hou:2017a} and \citet{Aoyama:2018a} showed that the dust abundance
is dominated by large ($\sim 0.1~\micron$) grains in the CGM
in their cosmological hydrodynamical simulations. However, their models failed to explain the reddening observed for a sample of Mg \textsc{ii} absorbers, which are considered to trace the CGM \citep{Menard:2012a}. This implies that our understanding of dust processing in the CGM is imperfect. \citet{Hirashita:2021a} showed that, if we consider turbulent cool clumps in the CGM, formation of small grains by shattering could occur efficiently. Indeed, some observations showed that the cool medium in the CGM is turbulent \citep{Qu:2022a,Chen:2023a}. Thus, the grains can be accelerated by grain--gas coupling to obtain velocities larger than the shattering threshold (typically of the order of $\sim 1$ km s$^{-1}$; \citealt{Jones:1996a}).
Sputtering also affects the grain size distribution by destroying small grains efficiently \citep{Tsai:1995a,Hirashita:2015a}.
Because various processes affect the grain size differently, it may be useful to predict the grain size distribution in the CGM in future work.

We expect that some ongoing and future sensitive observations are useful in further constraining our models. Indeed, the \textit{James Webb Space Telescope} is sensitive enough to directly image mid-infrared dust emission in the CGM; thus, we will report some comparison studies in future work. Also, the \textit{Nancy Grace Roman Space Telescope} could be used to derive the reddening curve (as done by \citealt{Menard:2010a} using the SDSS data) with its multi-wavelength coverage.\footnote{\url{https://roman.gsfc.nasa.gov/science/WFI_technical.html}} for a larger sample of background QSOs than SDSS.

\section{Conclusion}\label{sec:conclusion}

To understand the origin and evolution of the CGM dust, we construct
a model of dust enrichment.
We treat the galaxy and the CGM as two distinct zones but consider mass exchange between them.
We assume that the medium is homogeneous within each zone.
In the galaxy, we apply our previous model, which is based on a chemical evolution model, for the mass evolution of
gas, metals, and dust.
We include stellar dust production, dust destruction in SN shocks, and dust growth by the accretion of gas-phase metals.
In the CGM, we consider dust supply from the galaxy via galactic outflows and dust destruction by sputtering.
We also incorporate inflows from the CGM to the galaxy under an assumption that the CGM acts as a
constant mass reservoir.

We parameterize the efficiencies or time-scales of principal processes
that could affect the dust abundance in the CGM, and
vary the following five parameters: the star formation time-scale in the galaxy ($\tau_\mathrm{SF}$), the infall time-scale from the CGM to the galaxy
($\tau_\mathrm{in}$), the dust condensation efficiency in stellar ejecta ($f_\mathrm{d\star}$), the mass loading factor for the outflow from
the galaxy to the CGM ($\eta_\mathrm{out}$), and the dust destruction time-scale by sputtering in the CGM ($\tau_\mathrm{dest}$).
We examine how these parameters affect the dust mass in the CGM.

Since the star formation activity in the galaxy directly affects
the metal/dust enrichment, $\tau_\mathrm{SF}$ affects the dust abundance
in the CGM.
The galactic outflow is the supply mechanism of dust to the CGM; thus,
$\eta_\mathrm{out}$ also strongly affects the CGM dust enrichment. 
If one of these two parameters ($\tau_\mathrm{SF}$ and $\eta_\mathrm{out}$) take
an extreme value, the dust enrichment in the CGM is inefficient for one of
the following reasons:
the gas in the galaxy is depleted too quickly (too large $\eta_\mathrm{out}$ or
too short $\tau_\mathrm{SF}$),
dust enrichment proceeds slowly in the galaxy (too long $\tau_\mathrm{SF}$), or
dust supply is inefficient (too small $\eta_\mathrm{out}$). Thus, we argue that there are optimum values for $\tau_\mathrm{SF}(\sim 5~\mathrm{Gyr}$) and $\eta_\mathrm{out}(\sim 1)$ that maximize the dust mass in the CGM.
The inflow regulated by $\tau_\mathrm{in}$ indirectly influences the dust mass in the CGM
since it fuels the star formation causing dust enrichment in the galaxy.
For the same reason as for $\tau_\mathrm{SF}$ and $\eta_\mathrm{out}$,
$\tau_\mathrm{in}$ has the `optimum' value ($\sim 3$ Gyr) that maximizes the dust abundance in the CGM.
The dust destruction time-scale $\tau_\mathrm{dest}$ has a straightforward influence on the dust mass in the CGM such that it monotonically
decreases as $\tau_\mathrm{dest}$ becomes short.
The dust condensation in stellar ejecta (i.e.\ $f_\mathrm{d\star}$) has little impact on the resulting dust mass in the CGM at later epochs ($t\gtrsim\tau_\mathrm{SF}$) when dust growth by accretion dominates the dust production.

We further compare the calculated dust mass with
an observationally derived dust mass in the CGM by \citet{Menard:2010a}.
Our predictions scaled with the typical stellar mass of the observational sample explains the observed dust mass well. Therefore, our model based on dust production in the galaxy and dust injection into the CGM is successful in explaining the dust abundance in the CGM.
The success of our simple model contributes to clarifying important processes for
dust enrichment in the CGM, and provides a basis on which future more complicated calculations such as cosmological hydrodynamical simulations are interpreted.

\section*{Acknowledgements}

We are grateful to the anonymous referee for
positive and useful comments, and to H. Inami for continuous support.
HH thanks the Ministry of Science and Technology (MOST) for support through a grant
MOST 108-2112-M-001-007-MY3, and the Academia Sinica
for Investigator Award AS-IA-109-M02.
MO thanks the Institute of Astronomy and Astrophysics of
Academia Sinica for the opportunity
of the Summer Students Program and the hospitality. 

\section*{Data Availability}

Data related to this publication and its figures are available on request from the corresponding author.

\bibliographystyle{mnras}
\bibliography{reference}

\bsp	
\label{lastpage}
\end{document}